\documentclass[12pt]{article}
\usepackage{amsmath}
\usepackage{latexsym}
\usepackage{bbm}
\usepackage{amssymb}

\def\a{\alpha}

\def\b{\beta}
\def\g{\gamma}

\def\d{\delta}

\def\e{\varepsilon}

\def\m{\mu}

\def\s{\sigma}
\def\t{\tau}

\def\G{\Gamma}

\def\pa{\partial}

\def\and{{\rm and}}

\def\ie{{\it i.e.,} }

\def\IZ{{\mathbbm Z}}

\def\and{{\quad {\rm and} \quad}}

\def\A5{{AdS_5 \times S^5}}

\newcommand{\be}{\begin{equation}}
\newcommand{\bea}{\begin{eqnarray}}

\newcommand{\ee}{\end{equation}}
\newcommand{\eea}{\end{eqnarray}}

\begin{document}
\vspace*{-1.0in}
\thispagestyle{empty}
\begin{flushright}
CALT-TH-2024-037
\end{flushright}

\normalsize
\baselineskip = 18pt
\parskip = 6pt

\vspace{1.0in}

{\Large \begin{center}
{\bf From Hadrons to Gravitons via Strings}
\end{center}}

\vspace{.25in}

\begin{center}
John H. Schwarz\footnote{jhs@theory.caltech.edu}
\\
\emph{Walter Burke Institute for Theoretical Physics\\
California Institute of Technology\\ Pasadena, CA  91125, USA}
\end{center}

\vspace{.25in}

\begin{center}
\textbf{Abstract}
\end{center}
\begin{quotation}
The first quantum string theories were developed around 1970, prior to 
the discovery of QCD, with the goal of producing a theory of hadrons.
Basic physical requirements and mathematical consistency of the string theories known at that time
turned out to require the inclusion of gravity and the existence of extra spatial dimensions. 
This came as a complete surprise to everyone who was involved. It
led to a completely new and very ambitious goal for string theory research,
namely a unified quantum theory of gravity and all other forces. In particular, this goal requires
that the string tension is 20 orders of magnitude larger than was previously envisioned.
Fifty years later, this goal is widely shared. 
\end{quotation}

\newpage


\pagenumbering{arabic}

\tableofcontents

\newpage

\section{Introduction}

String theory was developed in the search for a theory
of the {\em strong nuclear force}.  In the 1960s, when the
search began, the only force for which there already was a
satisfactory quantum theory was the {\em electromagnetic
force}. Quantum electrodynamics (QED) was, and still is, a well-established quantum field
theory of electrons and photons.
The study of gravity was a very remote, and seemingly irrelevant, topic
for people interested in particle physics. 

The goal of the high-energy theory group in Berkeley in the 1960s was to construct
the S matrix that encodes the scattering amplitudes of the strongly interacting particles,
which are called hadrons.
The correct theory of hadrons, quantum chromodynamics (QCD), was formulated in 1973.
It is a Yang--Mills theory of interacting {\em quarks} and {\em gluons}, 
the fundamental constituents of hadrons, that is based on the Yang--Mills gauge group $SU(3)$.
It is fortunate that QCD wasn't discovered a few years earlier!
Had this happened, it is unlikely that string theory would have been developed --
at least for a very long time.

The leaders of S-matrix theory research in Berkeley in the 1960s
were Geoffrey Chew and Stanley Mandelstam. Chew argued that
quantum field theory would not be useful for the strong interactions, since the
expansion parameter is too large. (This turned out to be wrong, at least in 
certain limits.) 
Instead, Chew and Frautschi \cite{Chew:1961ev} proposed that one could deduce the hadronic
S matrix, which encodes all of the scattering amplitudes, from some general principles:
\begin{itemize}
\item Unitarity and analyticity of the S matrix
\item Analyticity in angular momentum (Regge Pole Theory)
\item The bootstrap hypothesis: particle exchanges produce the forces that 
are responsible for their own existence.
\end{itemize}

Regge had realized previously that it is
sometimes useful to interpolate between the discrete physical values 
of angular momentum and consider continuous values.
Hadrons (especially baryons, many of which were discovered in the 1960s)
were observed to lie on approximately linear and parallel {\em Regge trajectories}
$$J = \a(s) = \a(0) + \a' s.$$
Whenever the function $\a(s)$ is a nonnegative integer $J$
(or half-odd integer in the case of fermions), there is a stable particle or unstable
resonance of spin $J$ and mass $M$, where $s = M^2$.
In the case of hadrons, the {\em Regge slope} $\a'$ is about $1.0\, {\rm GeV}^{-2}$.

In a reaction $1 + 2 \to 3 + 4$, energy and momentum are conserved
$$p_1^\m + p_2^\m = p_3^\m + p_4^\m ,$$
where $\mu =0,1,2,3$ for four-dimensional spacetime. The quantities
$$ s=(p_1 + p_2)^2, \quad t =(p_1 -p_3)^2, \quad u = (p_1 - p_4)^2$$
are computed using the Lorentz metric.
$s + t + u$ is the sum of the squares of the four masses. The bootstrap conjecture is that
particle exchanges provide the forces that are responsible for their own existence.
This conjecture implies that in a first approximation that neglects resonance widths, scattering amplitudes are given by formulas of the sort
$$A(s,t) = \sum_i \frac{\b_i(t)}{s - M_i^2}
= \sum_i \frac{\b_i(s)}{t - M_i^2} ,$$
and the residues $\b_i$ are polynomials that are determined by the spins that contribute.
Such a formula is only conceivable if there is an infinite spectrum of particles. 
Then these series are defined outside their regions of convergence by analytic continuation. 
This is different from conventional quantum field theory, with a finite spectrum,
where the two expressions would correspond to distinct Feynman diagrams that should be added together.

\section{The origins of string theory}

In 1968 Veneziano found an explicit realization of the bootstrap and Regge
behavior in the narrow-resonance approximation \cite{Veneziano:1968yb}. 
A somewhat simplified version of his formula is
$$ A(s,t) = g^2 \, B(-\a(s), - \a(t)),$$
where $B$ is Euler's beta function
$$B(x,y) = \frac{\G(x) \G(y)}{\G(x+y) }.$$
The Regge trajectories are linear $\a(s) = \a(0) + \a' s$.
This formula satisfies the bootstrap equations! It is much simpler than anyone
previously thought could be possible.
Soon thereafter Virasoro proposed, as an alternative,
$$  g^2 \frac{\G(- \frac{\a(s)}{2}) \G(- \frac{\a(t)}{2})
\G(- \frac{\a(u)}{2})}{\G(-\frac{\a(t) +\a(u)}{2})
\G(-\frac{\a(s) +\a(u)}{2})\G(-\frac{\a(s) +\a(t)}{2}) },$$
which has similar virtues \cite{Virasoro:1969me}.
Remarkably, these two formulas, guessed without a theory, later turned out to be, almost precisely,
tree-approximation amplitudes for theories of open-string and closed-string
scattering respectively! In the modern interpretation, discussed later, 
they correspond to gauge theory and gravity interactions.

The discussion that follows
is only intended to give an impressionistic view, omitting various technical details.
The $N$-particle generalization of the Veneziano
formula (found in 1969 by several groups) is
$$ 
A_N = g^{N-2}\int  \m_N(y)
\prod_{j<k} (y_j - y_k)^{\a' p_j\cdot p_k}\prod_{i=1}^N dy_i . 
$$
The measure $\m_N(y)$ contains step functions that ensure
that $y_{j+1} \geq y_j$. It also contains delta functions such that there are only
$N-3$ integrations. Altogether, $A_N$ has cyclic symmetry in the $N$ particles.
By a change of variables $A_N$ can be brought to a form in which the
$N$ particles are at points on the boundary of a circular disk (instead of on an
infinite line) with a specified cyclic ordering. 
The particles in this formula are required to belong to the adjoint
representation of a gauge group such as $SO(n)$ or $Sp(n)$.
The complete tree-approximation amplitude then takes the form
$$T_N = \sum_{perms} C_N A_N.$$
The sum is over all inequivalent cyclic orderings of the $N$ particles.
The group-theoretic coefficients, $C_N$, are traces
of a product of matrices associated to the particles
with the corresponding cyclic ordering. 

Shapiro's $N$-particle generalization of the Virasoro formula
is an analogous formula involving $N-3$ complex integrations \cite{Shapiro:1970gy}
$$ 
T_N = g^{N-2}\int \tilde\m_N(z) \prod_{i<j} |z_i - z_j|^{\a' k_i
\cdot k_j}\prod_{i=1}^N d^2 z_i .
$$
This formula has total symmetry in the $N$ particles, and
there is no associated gauge symmetry group. The coordinates $z_i$ can be recast
as $N$ points on a two-dimensional sphere.

These formulas have consistent
factorizations on well-defined spectra of single-particle states
\cite{Fubini:1969qb}\cite{Fubini:1969wp}, which can be incorporated 
in a Fock space generated by an infinite set of harmonic oscillators.
There is one such set of oscillators in the Veneziano case and two sets in
the Shapiro--Virasoro case. 
Consistent factorization was the first indication that these formulas
arise from {\em theories}, not just intriguing formulas.

In 1970 Nambu, Nielsen, and Susskind independently
interpreted the spectrum and amplitudes as arising from
a theory of a one-dimensional structure, later called a {\em string}:
open strings (with two ends) in the Veneziano case;
closed strings (or loops) in the Shapiro--Virasoro case.
Remarkably, the formulas preceded the interpretation!
The string tension is $T = 1/(2\pi \a')$.
In a Feynman diagram perspective, strings sweep out two-dimensional {\em world-sheets}
in spacetime. Having found the spectrum and tree-approximation amplitudes,
it became possible to study radiative corrections (loop amplitudes).
These are given by integrals associated to higher-genus
Riemann surfaces. These surfaces have boundaries when particles
associated to open strings are involved.

In 1970 Gross et al. calculated four-particle open-string one-loop amplitudes for which the string
world sheet is topologically a cylinder, \ie it has two circular
boundaries \cite{Gross:1970eg}. There were two cases of interest.
In the first case, all four particles are attached to one
of the boundaries, and in the second case two particles are attached to each boundary
(particles 1 and 2 attach to one boundary, and particles 3 and 4 attach to
the other boundary.) In the first case the amplitude is given by an integral 
that has a divergence. It can be removed, leaving a 
satisfactory finite result, by a procedure previously
introduced by Neveu and Scherk \cite{Neveu:1970iq}. So it is okay.
The second case (two particles on each boundary) turned out to be more
profound. The amplitude contains
{\em branch points} in the variable $s = (p_1 + p_2)^2$ that {\em violate unitarity}.
Unless one could eliminate these branch points, this could not be a consistent quantum theory.
This was a serious problem!

Lovelace rescued the theory by
observing that these singularities would become poles (rather than branch points) if
$$
\a(0)=1 \quad {\rm and} \quad d=26,
$$
where $d$ is the dimension of spacetime 
($d-1$ spatial dimensions and one time dimension) \cite{Lovelace:1971fa}.
Nobody working in this field had questioned the assumption that $d=4$ before this.
Lovelace's discovery forced us to take {\em extra dimensions
of space} (22 of them in this case) seriously for the first time. 
We were aware of the work of Kaluza and Klein many years earlier, 
but that was in the context of gravity.
Why should a theory of hadrons require extra dimensions of space? 
This was very surprising and unexpected. $\a(0)=1$ implies that the spectrum contains
massless spin one particles and spin 0 tachyons, so these are
additional issues. Lovelace's poles describe
closed-string intermediate states in the reaction $1 + 2 \to 3 + 4$. This
could have been anticipated, because the cylinder can be
viewed either as a one-loop open-string diagram or finite propagation of a closed string! 
Thus, Lovelace's poles correspond to closed-string states. This
was the discovery of {\em open-string -- closed-string duality}.

The open-string particle spectrum of the $d=26$ critical string theory 
has an infinite algebra of constraints generated by operators $L_m$, $m\in \IZ$ that satisfy the 
Virasoro algebra \cite{Virasoro:1969zu}
$$
[L_m, L_n] = (m-n) L_{m+n} + \frac{c}{12} (m^3 - m)\d_{m+n, 0}.
$$
$c=d=26$ for the critical bosonic string. These constraints eliminate unphysical negative-norm states
(called ghosts) from the physical spectrum.
The Virasoro operators generate the group of conformal symmetries of the two-dimensional
string world-sheet theory. There is a similar story for the closed-string spectrum involving
two copies of the Virasoro algebra.

In 1971 Ramond introduced a string theory
analog of the Dirac equation \cite{Ramond:1971gb}, which is the wave equation for a free electron. His
proposal was that just as the string's momentum $p^\m$ is the zero mode of a
string momentum density $P^\m (\s)$, the Dirac matrices $\g^\m$ should be the
zero modes of densities $\G^\m (\s)$. Then he defined Fourier modes
$$ 
F_n = \frac{1}{2\pi}\int_0^{2\pi} e^{-in\s} \G \cdot P d\s \qquad n\in \IZ.
$$
In particular, $ F_0 = \g \cdot p + {\rm oscillator \, terms}$. He then proposed the wave equation
$$ 
(F_0 + M) |\psi\rangle =0,
$$
which is a stringy generalization of the Dirac equation. (We later realized that the 
interacting theory containing these fermions requires $M=0$.)
Ramond also observed that the Virasoro algebra generalizes to
a super-Virasoro algebra with odd elements $F_n$ and even elements $L_n$.
$\{F_m, F_n\} = F_m F_n + F_n F_m = L_{m+n}$, aside from a central extension, etc. 
Algebras that include anticommutators
in addition to commutators are called {\em superalgebras}. This motivated
the mathematician Kac to give a complete classification of simple 
superalgebras \cite{Kac:1977em}, which constitutes an extension of
Cartan's classification of ordinary (bosonic) Lie algebras.


Very soon after Ramond's work, Neveu and I introduced a second bosonic 
string theory \cite{Neveu:1971rx}.
It involves a similar operator to Ramond's, but the periodic density 
$\G^\m (\s + 2\pi) =\G^\m(\s)$ is
replaced by an antiperiodic one $H^\mu (\s + 2\pi) = -H^\mu (\s)$. In terms of modes
$$ 
G_r = \frac{1}{2\pi}\int_0^{2\pi} e^{-ir\s} H \cdot P d\s \qquad r \in \IZ + 1/2
$$
are the odd elements of a super-Virasoro algebra very similar to the one found by Ramond. 
These bosons and Ramond's fermions combine into a unified theory of bosons and fermions.
This was an early version of what would later become superstring theory. Additional important
facts remained to be understood first. The titles of these papers show that we were 
still thinking about hadrons. The critical spacetime dimension of the RNS string is $d=10$. 
At the time, we considered this to be a step in the right direction, 
and we hoped that the next theory would have $d=4$.

Later in 1971 Gervais and Sakita constructed a world-sheet action for the 
RNS string \cite{Gervais:1971ji}
$$ 
S = T\int d\s d\t \left( \pa_\a X^\m \pa^\a X_\m -i \bar\psi^\m \g^\a 
\pa_\a \psi_\m\right) .
$$
They pointed out that this simple theory has global supersymmetry
$$
\d X^\m = \bar\e \psi^\m, \quad \quad \d\psi^\m = -i \g^\a \e \pa_\a X^\m ,
$$
where $\e$ is an infinitesimal constant Grassmann-valued spinor.
Even though superalgebras had already been discussed, this is the very first theory 
ever shown to have {\em supersymmetry}. It is a free theory in two dimensions, 
which is about as simple as possible. The symmetry relates
bosonic operators $X^{\mu}$ and fermionic operators $\psi^{\mu}$.
This was previously believed not to be possible \cite{Coleman:1967ad}. 
Five years later the Gervais--Sakita global supersymmetry together with the super-Virasoro
constraints were shown to result from gauge fixing an action with {\em local} 2d supersymmetry
\cite{Deser:1976rb}\cite{Brink:1976sc}. 

The Gervais--Sakita discovery motivated Wess
and Zumino to construct four-dimensional interacting analogs \cite{Wess:1973kz}\cite{Wess:1974tw}.
This then inspired the construction of supersymmetric extensions of the standard model
and subsequent experimental searches for supersymmetry partners. Even though relatively
few physicists cared about strings in this era, many became interested in supersymmetry
as possible new physics to be discovered at accelerators.


\section{The demise of string theory}

In 1973--74 there were many good reasons to stop working on string
theory: a successful and convincing theory of hadrons (QCD) was
discovered, and string theory had severe problems as a theory of hadrons.
These included an unrealistic spacetime dimension ($d=10$ or $d=26$),
an unrealistic spectrum (including tachyons and massless particles),
and the absence of point-like constituents, such as quarks and gluons. 
A few years of attempts to do better had been unsuccessful.
The success of QCD eliminated the need to formulate a
theory of hadrons based on strings. 

Moreover, convincing theoretical and experimental evidence for the
Standard Model was rapidly falling into place. That was where the
action was. Even for those seeking to pursue speculative theoretical ideas
there were options other than string theory
that most of them found more appealing, such as
grand unification and supersymmetric field theory. Understandably,
string theory fell out of favor. What had been a booming enterprise
involving several hundred theorists rapidly came to a grinding halt.
Only a few diehards continued to pursue it. Even today,
it remains an open question whether there exists a string theory,
not yet discovered, that is
equivalent to QCD. Such a dual description of QCD could be useful.

\section{Gravity and unification}

Yoneya interpreted the massless spin 2 state
in the closed-string spectrum as a {\em graviton} \cite{Yoneya:1973ca}\cite{Yoneya:1974jg}.
With this identification and a theorem of Weinberg,
it was easy to show that this string theory particle has the same interactions as
the graviton in general relativity at low energies (compared to the string scale).
Similarly, the massless spin 1 states in the open-string spectrum could be
interpreted as gauge-theory particle \cite{Neveu:1971mu}.

A few months later, unaware of Yoneya's prior work, 
Scherk and I rediscovered the graviton in the string spectrum.
This led us to propose interpreting string theory
as a {\em quantum theory of gravity, unified with gauge theory forces} 
rather than as a theory of hadrons \cite{Scherk:1974ca}.
This requires that the string length scale is roughly the Planck scale ($10^{-33}$ cm)
rather than the nuclear scale ($10^{-13}$ cm). So the size of the strings decreased
by 20 orders of magnitude, and their tensions increased by the same factor.
This proposal had several advantages:

\noindent $\bullet$ The existence of gravity is {\em predicted} by the theory.

\noindent $\bullet$ String theory has no UV divergences.

\noindent $\bullet$ In a gravitational theory extra dimensions could be a good thing.
The 4d effective theory is determined by the details of the geometry of the
compact extra dimensions, which are determined dynamically.

\noindent $\bullet$ Gravity is unified with gauge theory forces.

This was a {\em serendipitous theoretical discovery} -- something that is very unusual.
The best known serendipitous scientific discoveries are experimental or observational.
A few examples are Dynamite: Nobel (1833); Insulin: Minkowski and von Mering (1889);
X-rays: Roentgen (1895); Radioactivity: Becquerel (1896); Penicillin: Fleming (1928);
Big Bang CMB: Penzias and Wilson (1965).

How was this proposal received?
Scherk and I spoke about our ideas at various conferences and seminars.
Everyone we spoke with was polite and showed interest. A few prominent physicists, 
such as Gell-Mann and Zumino, and later Witten,
said that this proposal was potentially very important.
Yet, it was largely ignored. The explanation lies in the sociology of the profession at that time.
{\em Relativists}, who thought about black holes, gravitational waves,
the geometry of the universe, etc.
had no use for a quantum theory of gravity. It was far removed from anything of interest to them.
{\em Particle theorists}, who were interested in understanding phenomena that
could be observed in current or future accelerators, had no interest in
gravity, which is very far out of reach. 
There was a lot of interest in supersymmetry, but not in gravity.
Both groups of scientists were correct in their opinions.
Therefore the two communities were completely disjoint. A proposal that
would bridge the gap had very few takers. 

\section{Supergravity and supersymmetric strings}

Supergravity is a supersymmetric extension of general relativity. It
was formulated for $N=1$ supersymmetry in four-dimensional spacetime in 1976 by
Freedman, van Nieuwenhuizen, and Ferrara \cite{Freedman:1976xh}. Very soon
thereafter Deser and Zumino \cite{Deser:1976eh}
introduced a clever way to simplify some of the calculations. This work was
very interesting and quickly received a lot of attention.

Also in 1976, Gliozzi, Scherk, and Olive proposed a
projection of the RNS string spectrum (both for bosons and fermions) -- the
{\em GSO Projection} -- that removes roughly half of the states in the RNS string 
spectrum including the tachyon \cite{Gliozzi:1976qd}. They showed that after 
the projection the number of bosons and fermions is equal at every mass level.
This was compelling evidence for {\em 10d spacetime
supersymmetry} of the GSO-projected theory, but it was not a proof.
Supersymmetry is necessary for consistency, because the string spectrum contains
a massless gravitino. The 1976 10d spacetime supersymmetry
proposed by GSO is completely different from the 1971 2d
world-sheet supersymmetry identified by Gervais and Sakita.
After the GSO projection the RNS theory has supersymmetry in 10d Minkowski spacetime. 
This symmetry can be spontaneously broken when extra dimensions are compactified. 

Also in 1976, Brink, Scherk, and I constructed supersymmetric Yang--Mills
theory in ten dimensions \cite{Brink:1976bc}. 
The spinor supercharge in 10d has 16 components, which is
the maximum possible for a Yang--Mills theory.
By itself, this 10d super Yang--Mills theory is only a classical field theory,
not a quantum theory, since it has bad UV divergences.
{\em Dimensional reduction} 
of the 10d theory was used to deduce the Lagrangian for $N =4$
super Yang--Mills theory in 4d spacetime for any Yang--Mills gauge group. 
It was later understood that these are conformally
invariant quantum field theories, which implies UV finiteness.
These theories have four 4-component
Poincar\'e supercharges and four 4-component conformal supercharges.
Though not realistic, these theories have played
a prominent role in research since 1997 in the context of 
AdS/CFT \cite{Maldacena:1997re}.

In 1978 Cremmer, Julia, and Scherk constructed $d=11$ supergravity \cite{Cremmer:1978km}. 
Eleven is the highest dimension possible for supergravity.
This theory contains three fields. In addition to graviton
and gravitino fields, there is a three-form tensor field.
In the early 1980s there was quite a bit of research studying
ways to compactify the seven extra dimensions in an attempt to obtain something 
realistic in four dimensions. These were useful exercises even though nothing realistic was found.
11d supergravity is very beautiful, but it has severe UV divergences.
A good question is whether it could be the low-energy approximation
to a consistent well-defined quantum theory.  In the mid 1990s it
became clear that this is the case, and Witten gave
the quantum theory the name {\em M theory}.

In 1979 Michael Green and I began a collaboration with the initial goal of
understanding the ten-dimensional spacetime supersymmetry of the GSO-projected
RNS string theory. Whenever possible, we also collaborated with Lars Brink.
Over the subsequent five years we formulated (and named) the type I,
type IIA, and type IIB superstring theories; proved the 10d spacetime supersymmetry of the
spectrum and interactions in each case; computed various tree and
one-loop amplitudes and elucidated their properties; formulated superstring field theory in the
light-cone gauge for the type I and type IIB theories;
formulated an alternative superstring world-sheet theory with
manifest 10d spacetime supersymmetry \cite{Green:1980zg}\cite{Green:1983wt}\cite{Green:1981yb}.

In 1983 I constructed the equations of motion of
the type IIB superstring in the low-energy supergravity approximation \cite{Schwarz:1983qr}.
The last page points out that the equations of motion have a solution
describing a 10d geometry of the form
$AdS_5 \times S^5$, which is analogous to the $AdS_4 \times S^7$
solution of 11d supergravity discovered 
a few years earlier by Freund and Rubin \cite{Freund:1980xh}.
These geometries feature in AdS/CFT. 
In particular, the $AdS_5 \times S^5$ solution
of type IIB superstring theory is dual to $N =4$ 4d super Yang--Mills theory.
The symmetry of both of them is given by the superalgebra $PSU(2,2|4)$. 

\section{Anomalies}

It is essential for the consistency of a quantum theory that local
symmetries (such as gauge symmetries and local Lorentz invariance) of
the tree-level/classical theory are not broken by quantum corrections.
Such symmetry-destroying quantum corrections, called {\em anomalies},
potentially occur at the one-loop level in parity-violating theories.
The standard model is an excellent example of a theory in which
various such anomalies beautifully cancel. If one were to ignore the quarks
or the leptons, the quantum theory would be inconsistent. When both
are included, however, their anomaly contributions cancel.

In 1984 we knew three superstring theories: Type I, Type IIA, and Type IIB.
Type IIA is parity conserving, and therefore it is anomaly-free. 
The type IIB theory is parity violating. In 1983 Alvarez--Gaum\'e and Witten
proved that all of the gravitational anomalies of the type IIB theory cancel \cite{Alvarez-Gaume:1983ihn}.
For these reasons, Green and I focussed our attention on potential anomalies of Type I superstring theory.
Classically, it is a parity-violating theory that is defined for any orthogonal or
symplectic gauge group.
There are two world-sheet topologies that contribute to pure-gauge anomalies: a cylinder
and a M\"obius strip. Each of them contributes the same structure, but with different coefficients.
We found that the two contributions only cancel if the gauge group is $SO(32)$ \cite{Green:1984qs}. 
Any other choice of orthogonal or symplectic group is inconsistent at the quantum level.

By an analysis of the low-energy effective field theory
we could also analyze local gravitational and mixed anomalies \cite{Green:1984sg}. They all
cancelled beautifully for $SO(32)$. (More precisely, the Lie group is $Spin(32)/\IZ_2$.)
To our surprise, we also discovered that the anomalies could cancel for a second gauge group,
namely $E_8 \times E_8$. Both of these groups are 496-dimensional and have rank equal to 16.
Also, their weight lattices are the two even self-dual lattices that exist in 16 dimensions.
We did not know a superstring theory with gauge group $E_8 \times E_8$, but it was
plausible that one should exist, and we set out to find it. 
Gross, Harvey, Martinec, and Rohm beat us to it. They
introduced the {\em heterotic string} for both gauge groups \cite{Gross:1984dd}.
Soon thereafter Candelas, Horowitz, Strominger, and Witten
introduced {\em Calabi--Yau compactification} of the six extra dimensions \cite{Candelas:1985en},
which leads to 4d effective theories with $N=1$ supersymmetry.
Applied to the $E_8\times E_8$ heterotic string theory, they showed that this could give rise to
many realistic features (and some unrealistic ones).

\section{Conclusion}

The history of string theory is one of {\em unification}: of particles and forces,
of particle theorists and relativists, of math and physics.
It has been an exciting journey that is still going strong.
After a decade in the shadows (1974-84), superstring theory
suddenly became a mainstream activity. 
A great deal has happened since 1984, a period of 40 years!
During the {\em second superstring revolution} in the mid 1990s
many additional important results were discovered:
dualities, M theory, black-hole entropy, F theory, etc.
Also, the AdS/CFT holographic duality
discovery was transformative. Current research directions
include new types of symmetries, the swampland program, and
celestial holography.

An obvious question is ``Where is the experimental evidence?"
Before the LHC turned on there was optimism about discovering new particles
that are supersymmetry partners of known particles, but so far that
has not happened. Such a discovery would not prove that 
string theory is correct, but it would be extremely informative,
perhaps leading eventually to a new standard model. Such a theory could make
a better target for ``top-down" approaches to aim for.
The recent version of a dark-dimension proposal suggests the
possible existence of a micron-scale fifth dimension that only supports gravitational strength
forces \cite{Montero:2022prj}. This might be accessible to experiment. 
Also, there is considerable effort studying the possible implications of string theory
for early universe cosmology. This might also be informative. It seems quite safe to predict
that it will take a very long time to figure out how to connect string theory to
experiment at all scales. Yet, I am optimistic that it is possible.

\section*{Acknowledgments}

This material is based upon work supported by the U.S. Department of Energy,
Office of Science, Office of High Energy Physics, under Award Number DE-SC0011632.

\end{document}